\documentclass[12pt,british]{article}
\usepackage[T1]{fontenc}
\usepackage[latin9]{inputenc}
\usepackage{geometry}
\usepackage{color}
\usepackage{array}
\usepackage{float}
\usepackage{multirow}
\usepackage{amstext}
\usepackage{amssymb}
\usepackage{setspace}
\usepackage[authoryear, round]{natbib}
\usepackage[hidelinks]{hyperref}
\makeatletter

\providecommand{\tabularnewline}{\\}

\date{}

\newtheorem{theorem}{Theorem}
\newtheorem{remark}{Remark}

\makeatother

\usepackage{babel}

\begin{document}

\title{Doubly robust goodness-of-fit test of coarse structural nested mean
models with application to initiating combination antiretroviral treatment
in HIV-positive patients}

\author{Shu Yang and Judith J. Lok}
\maketitle
\begin{abstract}
Coarse structural nested mean models provide a useful tool to estimate
treatment effects from longitudinal observational data with time-dependent
confounders. However there is no existing guidance to specify the
treatment effect model, and model misspecification can lead to biased
estimators, preventing valid inference. To test whether the treatment
effect model matches the data well, we derive a goodness-of-fit test
procedure based on overidentification restrictions tests. We show
that our test statistic is doubly-robust in the sense that with a
correct treatment effect model, the test statistic has the correct
level if either the treatment initiation model or a nuisance regression
outcome model is correctly specified. In a simulation study we show
the test procedure has correct type-I error and is powerful to detect
model misspecification. In addition, we apply the test procedure to
study how the timing of combination antiretroviral treatment initiation
after infection predicts the one year treatment effect in HIV-positive
patients with acute and early infection.

\end{abstract}

\section{Introduction}

The gold standard for evaluating effects of interventions are randomized
controlled trials. However, they are not always available; for example,
for evaluating a treatment strategy for HIV-positive patients, a randomized
controlled trial would force patients to take the treatment or to
be off the treatment regardless of their health status. Observational
studies are useful in these settings. In observational studies, there
often is time-dependent confounding by indication: some covariates
are predictors of both the subsequent treatment and outcome, and are
also affected by the past treatment history. Then, standard methods
adjusting for the covariates history are fallible and can lead to
bias (\citealp{robins1992g}; \citealp{robins2000marginal}; \citealt{robins2000marginalstructural}).

Coarse structural nested mean models (\citealp{robins1998correction})
provide a useful tool to estimate treatment effects from longitudinal
observational data. \citet{lok2012impact} developed a time-dependent
version of coarse structural nested mean models and applied it to
investigate the impact of the timing of combination antiretroviral
treatment initiation on the effect of one year treatment in HIV-positive
patients. Their semiparametric method leads to an infinite number
of unbiased estimating equations and a huge class of consistent and
asymptotically normal estimators. An optimal estimator can be derived
within this class of coarse structural nested mean models under well-specified
models for the treatment effect, treatment initiation, and a nuisance
regression outcome model, in an unpublished 2014 technical report
available from the second author. The key assumption lies in a well-specified
model for the treatment effect. However, no guidance exists on how
to specify the treatment effect model, and model misspecification
may lead to biased estimators, preventing valid inference. 

The main contribution of this article is to derive a goodness-of-fit
test statistic for testing correct specification of the treatment
effect model. The key insight is that with a correctly-specified treatment
effect model we have more unbiased estimating equations than the number
of parameters, which results in overidentification of the parameters.
Overidentification restrictions tests, also called Sargan tests or
$J$-tests (\citealp{sargan1958estimation} and \citealp{hansen1982large}),
are widely used in the econometric literature; they however seem to
have been previously unnoticed in the biostatistics literature. The
standard overidentification restrictions test, given by the minimized
value of the generalized method of moments (\citealp{newey1994large};
\citealp{imbens1995information}) criterion function, has a chi-squared
limiting distribution, with degrees of freedom equal to the number
of overidentification restrictions. In most situations, the minimum
of the generalized method of moments criterion is obtained by a continuous
iterative procedure to update the parameter estimates until convergence
\citep{hansen1996finite}. \citet{arellano1991some} showed the test
statistic based on one-step estimates other than the optimal generalized
method of moments estimates is not robust and tends to over-reject
even in large samples. Our test procedure is different from  the standard
overidentification restrictions tests in this regard. We do not obtain
parameter estimates by minimizing an objective function, but rather
we obtain parameter estimates by solving  the optimal estimating equations
with the number of equations equal to the number of parameters. The
overly identified restrictions are only used for testing, not for
estimation. This difference allows us to greatly reduce the computation
burden. Our simulation studies show that our test statistic has correct
size for large samples under the scenarios we considered. Another
merit of the overidentification restrictions test is that no bootstrap
is needed to compute the test statistic, which could be valuable with
the large samples that are increasingly common.

\section{Motivating problem and basic setup \label{sec:Motivating-Data}}

\subsection{The motivating problem}

Combination antiretroviral treatment is the standard initial treatment
for HIV, and has considerably reduced the morbidity and mortality
in HIV-positive patients. In the HIV literature, findings imply that
there are key early events, during acute and early infection, in the
pathogenesis of HIV infection that determine the long-term pace of
disease progression \citep{hecht2006multicenter}. However, there
is no strong evidence to support when to start treatment in patients
in the acute and early stages of infection. It is important to understand
the effect of initiating treatment at different times during the course
of HIV infection. This investigation relies on an observational study,
where we emulate a counterfactual experiment using causal models.

\subsection{The Acute Infection and Early Disease Research Program}

The Acute Infection and Early Disease Research Program study is a
multicenter, observational cohort study of $1762$ HIV-positive patients
diagnosed during acute and early infection (\citealp{hecht2006multicenter}).
Dates of infection were estimated based on a stepwise algorithm that
uses clinical and laboratory data (\citealp{smith2006lack}). We included
patients with CD4 and viral load measured within $12$ months of the
estimated date of infection, which resulted in $1696$ patients. Let
$m$ denote the number of months between the estimated date of infection
and combination antiretroviral treatment initiation ($m=0,\ldots,11$),
where $0$ indicates the estimated date of infection. We are interested
in evaluating the impact of $m$ on the effect of one year treatment.

\subsection{Notation}

Let $Y_{k}$ be the patient's CD4 count at month $k$ since the estimated
date of infection $(k=0,\ldots,K+1\equiv24)$, and $L_{m}$ be a vector
of covariates measured at month $m$, including age, gender, race,
injection drug use, CD4 count and viral load. Let $A_{m}$ be one
if the patient was on treatment at month $m$ and zero otherwise.
We assume that once treatment is started, it is never discontinued.
We use overbars to denote a variable history; for example, $\bar{A}_{m}$
is the treatment history until month $m$. Let $T$ be the actual
treatment initiation time. The patients are assumed to be an independent
sample from a larger population\textcolor{black}{{} \citep{rubin1978bayesian},
and for notational simplicity we drop the subscript $i$ for patients.
To handle missingness, for $L$, if this is missing at month $m$,
$L_{m}$ was coded as ``missing''. For intermediate missing outcomes,
we imputed $Y_{k}$ by interpolation; if the outcome is missing just
prior to onset of treatment, we imputed $Y_{k}$ by carrying the last
observation forward. }Let $X\equiv(\bar{A}_{K},\bar{L}_{K},\bar{Y}_{K+1})$
denote the patient's full record.\textcolor{black}{{} Until Section
6 we assume all patients are followed up until month $K+1$.} 

\textcolor{black}{Let $Y_{k}^{(m)}$ be the CD4 count at month $k$,
possibly counterfactual, had the patient started treatment at month
$m$. Let $Y_{k}^{(\infty)}$ be the CD4 count at month $k$ had the
patient never started treatment during the course of follow up. We
assume the patient's observed outcome $Y_{k}$ is equal to the potential
outcome $Y_{k}^{(m)}$ for $m$ equal to the actual treatment initiation
time $T$; that is, if $k>T$,} $Y_{k}=Y_{k}^{(T)}$ and if $T\geq k$,
$Y_{k}=Y_{k}^{(\infty)}$. 

We assume the assumption of no unmeasured confounding (\citealp{robins1992g}):
\begin{equation}
Y_{k}^{(\infty)}\amalg A_{m}\mid\bar{L}_{m},\bar{A}_{m-1}\ (k=m+1,\ldots,m+12),\label{eq:NUC}
\end{equation}
where $\amalg$ denotes ``is independent of'' \citep{dawid1979conditional}.
This assumption holds if $\bar{L}_{m}$ contains all prognostic factors
for $Y_{k}^{(\infty)}$ that affect the treatment decision at month
$m$. For example, if patients with lower CD4 counts initiated treatment
earlier, the assumption (\ref{eq:NUC}) would fail to hold if $\bar{L}_{m}$
does not include the history of the CD4 count.

\subsection{Coarse structural nested mean model }

We model the treatment effect, comparing treatment starting at month
$m$ to never starting among the subgroup of patients with covariate
history $\bar{l}_{m}$ and $T=m$, as\textit{
\begin{equation}
E\{Y_{k}^{(m)}-Y_{k}^{(\infty)}\mid\bar{L}_{m}^{(\infty)}=\bar{l}_{m},T=m\}=\gamma_{m,\psi}^{k}(\bar{l}_{m})\ (k=m,\ldots,m+12),\label{eq:SNMM}
\end{equation}
}where $\psi$ is the parameter in the treatment effect model. From
now on, we consider $\gamma_{m,\psi}^{k}(\bar{l}_{m})=(\psi_{1}+\psi_{2}m)(k-m)1_{(m\leq k)},$
with $(k-m)$ the duration of treatment from month $m$ to month $k$.\textcolor{black}{{}
We restrict the range of $k$ from $12$ to $K+1$, whereby we avoid
making extra modeling assumptions beyond the necessary ones to estimate
$\gamma_{m}^{m+12}(\bar{l}_{m})$ in order to gain robustness. }Particularly,
$\gamma_{m,\psi}^{m+12}(\bar{l}_{m})$\textcolor{red}{{} }quantifies
the effect of one year treatment if HAART was initiated at month $m$,
among the subgroup of patients with covariate history $\bar{l}_{m}$.
If outcome is the CD4 count and $\gamma_{m,\psi}^{m+12}(\bar{l}_{m})>0,$
the effect of one year treatment is beneficial. $12\psi_{1}$ quantifies
the effect of one year treatment if treatment was started at the estimated
date of infection, and $\psi_{2}$ quantifies the increase of the
effect of treatment for each month of delay after the estimated date
of infection. Under this model, the treatment effect is homogeneous.\textcolor{blue}{{}
}In practice, the treatment effect may vary among different groups;
for example, male and female patients may have different responses
to the combination antiretroviral treatment. We can then extend the
model as $\gamma_{m,\psi}^{k}(\bar{l}_{m})=(\psi_{1}+\psi_{2}m+\psi_{3}\mathrm{gender})(k-m)1_{\{m\leq k\}},$
where $\psi_{3}$ quantifies the magnitude and direction of the impact
of gender.

For $k=12,\ldots,K+1$, define $H(k)=Y_{k}-\gamma_{T}^{k}(\bar{L}_{T})$.
As proved in \textcolor{black}{\citet{robins1992g}, \citet{lok2004estimating}
and \citet{lok2012impact}, $H(k)$ is mimicking a counterfactual
outcome $Y_{k}^{(\infty)}$ in the sense that for $k=12,\ldots,K+1$
and $m=k-12,\ldots,k-1$, 
\begin{equation}
E\{H(k)\mid\bar{L}_{m},\bar{A}_{m-1}=\bar{0},A_{m}\}=E\{Y_{k}^{(\infty)}\mid\bar{L}_{m},\bar{A}_{m-1}=\bar{0},A_{m}\},\label{eq:imp1ofNUC}
\end{equation}
since by subtracting from the observed $Y_{k}$ the average effect
of treatment, we would obtain the quantity that mimics the outcome
had the patient not been treated. The implication of (\ref{eq:imp1ofNUC})
and the assumption of no unmeasured confounding is that for $k=12,\ldots,K+1$
and $m=k-12,\ldots,k-1$, 
\begin{equation}
E\{H(k)\mid\bar{L}_{m},\bar{A}_{m-1}=\bar{0},A_{m}\}=E\{H(k)\mid\bar{L}_{m},\bar{A}_{m-1}=\bar{0}\},\label{eq:assumption}
\end{equation}
which plays a key role for estimation. }

\subsection{The conditional probability of treatment initiation}

We use a pooled logistic regression to model the probability of treatment
initiation at month $m$, conditional on the past history, $p_{\theta}(m)\equiv P(A_{m}=1\mid\bar{A}_{m-1}=\bar{0},\bar{L}_{m};\theta)=1_{(\bar{A}_{m-1}=\bar{0})}1_{\mathrm{visit}}(m)/[1+\exp\{-\theta^{T}f(\bar{L}_{m})\}],$
where $1_{\mathrm{visit}}(m)$ is an indicator of whether a visit
took place at month $m$, and $f(\bar{L}_{m})$ is some function of
$\bar{L}_{m}$. Let $J_{\mathrm{trt}(\theta)}(X)$ denote the estimating
function for $\theta_{0}$.

\subsection{The unbiased estimating equation and optimal estimation}

Model (\ref{eq:SNMM}) cannot be fit by standard regression methods
because it involves potential outcomes. However, one can get consistent
estimates by constructing unbiased estimating equations based on (\ref{eq:assumption})
\citep{lok2012impact}: for any measurable, bounded function $q_{m}^{k}:\bar{\mathcal{L}}_{m}\rightarrow\mathbb{R}^{p}$,
$k=12,\ldots,K+1$, $m=k-12,\ldots,k-1$, let 
\[
G_{(\psi,\theta,q)}(X)\equiv\sum_{k=12}^{K+1}\sum_{m=k-12}^{k-1}q_{m}^{k}(\bar{L}_{m})[H_{\psi}(k)-E\{H_{\psi}(k)|\bar{L}_{m},\bar{A}_{m-1}=\bar{0}\}]\{A_{m}-p_{\theta}(m)\}.
\]
We use empirical process notation throughout. We let $P$ denote the
probability measure induced by $X$ and let $P_{n}$ denote the empirical
measure induced by $X_{1},...,X_{n}$. Given a measurable function
$f:\mathcal{X}\mapsto\mathbb{R}$, we write $P_{n}f(X)=n^{-1}\sum_{i=1}^{n}f(X_{i})$
and $Pf(X)$ for the expectation under $P$. Then 
\begin{equation}
P_{n}\{\begin{array}{cc}
G_{(\psi,\theta,q)}(X)^{T} & \ \ J_{\mathrm{trt}(\theta)}\end{array}(X)^{T}\}^{T}=0\label{eq:ee}
\end{equation}
are the stacked unbiased estimating equations for both the parameter
$\psi$ and the (nuisance) parameter $\theta$. For simplicity, we
will suppress the dependence of the estimating functions on $X$;
for example, $P_{n}G_{(\psi,\theta,q)}$ is shorthand for $P_{n}G_{(\psi,\theta,q)}(X)$.
Sometimes, we also drop the dependence on the parameters. 

\textcolor{black}{In theory, $q$ can be chosen arbitrarily; however
it largely influences the precision of the resulting estimator. To
derive the optimal estimating equation, and therefore the optimal
estimator, we assume that }for $k,s=12,\ldots,K+1$ and $m$ with
$m=\max(k-12,s-12),\ldots,\min(k-1,s-1,K)$,
\begin{equation}
\mathrm{cov}\{H(k),H(s)\mid\bar{L}_{m},\bar{A}_{m-1}=\bar{0},A_{m}\}\amalg A_{m}\mid\bar{L}_{m},\bar{A}_{m-1}.\label{eq:Homo}
\end{equation}
This assumption is a natural extension of (\ref{eq:assumption}).
This would be true under \citet{robins1998structural}'s rank preservation
assumption and $(Y_{k}^{(\infty)},Y_{s}^{(\infty)})\amalg A_{m}\mid\bar{L}_{m},\bar{A}_{m-1}$.
However, the rank preservation assumption is unlikely to hold in practice
and the assumption (\ref{eq:Homo}) is weaker. 

The optimal estimating equations, within the class of $P_{n}G_{(\psi,\theta,q)}$
indexed by $q$ for any measurable and bounded functions $q_{m}^{k}$,
can be obtained by finding $q^{\mathrm{opt}}$ that satisfies $E\{\partial/\partial\psi^{T}G_{(\psi_{0},\theta_{0},q)}\}=E\{G_{(\psi_{0},\theta_{0},q)}G_{(\psi_{0},\theta_{0},q^{\mathrm{opt}})}^{T}\}$
for any $q$ \citep{newey1994large}. Then, under (\ref{eq:Homo})
\begin{eqnarray}
 &  & q_{m}^{\mathrm{opt}}(\bar{L}_{m})^{T}\equiv\{\mathrm{var}(H_{m}\mid\bar{L}_{m},\bar{A}_{m-1}=\bar{0})\}^{-1}\nonumber \\
 &  & \times\left\{ E\left(\frac{\partial}{\partial\psi}H_{m}\mid\bar{L}_{m},\bar{A}_{m-1}=\bar{0},A_{m}=1\right)-E\left(\frac{\partial}{\partial\psi}H_{m}\mid\bar{L}_{m},\bar{A}_{m}=\bar{0}\right)\right\} ,\label{eq:opt}
\end{eqnarray}
where $q_{m}^{\mathrm{opt}}=(q_{m}^{\mathrm{opt},l},\ldots,q_{m}^{\mathrm{opt},r})$
with $l=\max(m+1,12)$ and $r=\min(m+12,K+1)$, which are informed
by the fact that $m+1\leq k\leq m+12$ and $12\leq k\leq K+1$; $H_{m}=\{H_{\psi}(l),\ldots,H_{\psi}(r)\}^{T}$;
$\mathrm{var}(H_{m}\mid\bar{L}_{m},\bar{A}_{m-1}=\bar{0})$ is a matrix
with elements $\Gamma_{ks}^{m}\equiv\mathrm{cov}\{H_{\psi}(k),H_{\psi}(s)\mid\bar{L}_{m},\bar{A}_{m-1}=\bar{0}\}$;
and $E(\partial/\partial\psi H_{m}\mid\bar{L}_{m},\bar{A}_{m-1}=\bar{0},A_{m}=1)-E(\partial/\partial\psi H_{m}\mid\bar{L}_{m},\bar{A}_{m}=\bar{0})=(\Delta_{m}^{l},\ldots,\Delta_{m}^{r})^{T}$
with $\Delta_{m}^{k}\equiv E\{\partial/\partial\psi H_{\psi}(k)\mid\bar{L}_{m},\bar{A}_{m-1}=\bar{0},A_{m}\}-E\{\partial/\partial\psi H_{\psi}(k)\mid\bar{L}_{m},\bar{A}_{m-1}=\bar{0}\}$. 

\begin{remark}\label{rmk1}For the optimal estimating equations,
defined by (\ref{eq:ee}) and (\ref{eq:opt}), we address two issues.
One issue is that $E\{H_{\psi}(k)\mid\bar{L}_{m},\bar{A}_{m-1}=\bar{0}\}$
and $q^{\mathrm{opt}}$ depend on $\psi$.\textcolor{red}{{} }\textcolor{black}{Following
\citet{lok2012impact}, }we use a preliminary consistent estimate
$\hat{\psi}_{p}$ to replace $\psi_{0}$ in $E\{H_{\psi}(k)\mid\bar{L}_{m},\bar{A}_{m-1}=\bar{0}\}$
and $q^{\mathrm{opt}}$. The rationale for this replacement is that
the estimating equations are unbiased for any fixed value of $\hat{\psi}_{p}$
when $p_{\theta}(m)$ is correct. If $\gamma_{m,\psi}^{k}$ is linear
in $\psi$, $\Delta_{m}^{k}$ does not depend on $\psi$. One choice
of \textcolor{black}{$\hat{\psi}_{p}$ is the optimal estimator if
$q_{m}^{k}$ is only non-zero for $k=m+12$.} Another issue is that
$\Delta_{m}^{k}$ and $E\{H_{\psi}(k)\mid\bar{L}_{m},\bar{A}_{m-1}=\bar{0}\}$,
even with $\psi$ in lieu of $\psi_{p}$, depend on the true unknown
distribution. We will use parametric models to approximate these quantities.
Let $E\{H_{\psi_{p}}(k)\mid\bar{L}_{m},\bar{A}_{m-1}=\bar{0}\}$ be
parametrized by $\xi_{1}$; for example $E_{\xi_{1}}\{H_{\psi_{p}}(k)\mid\bar{L}_{m},\bar{A}_{m-1}=\bar{0}\}$
is a linear regression model with covariates $\bar{L}_{m}$. Likewise,
let $\Delta_{m}^{k}$ be parametrized by $\xi_{2}$. Denote estimating
functions for $\xi_{1}$, $\xi_{2}$ and $\psi_{p}$ by \textit{$J_{1(\xi_{1},\psi_{p})}$}\textcolor{black}{,}\textit{
$J_{2(\xi_{2})}$ }and\textit{ }$G_{p(\psi_{p},\xi_{2})}$.\textit{
}Since $G_{p}$ depends on $\Delta_{m}^{m+12}$, it is also a function
of $\xi_{2}$. 

\end{remark}

\section{Asymptotic results of optimal estimators\label{sec:Asymptotic-Results} }

We present the consistency and asymptotic normality result of the
optimal estimator. These results are the building blocks to derive
the goodness-of-fit test statistic. 

\begin{theorem}[Consistency] \label{Thm1} Let $G_{(\psi,\psi_{p},\xi,\theta)}^{*}$
be optimal estimating functions 
\[
G_{(\psi,\psi_{p},\xi,\theta)}^{*}=\sum_{k=12}^{K+1}\sum_{m=k-12}^{k-1}q_{m,\psi_{p},\xi_{2}}^{k,\mathrm{opt}}(\bar{L}_{m})[H_{\psi}(k)-E_{\xi_{1}}\{H_{\psi_{p}}(k)\mid\bar{L}_{m},\bar{A}_{m-1}=\bar{0}\}]\{A_{m}-p_{\theta}(m)\},
\]
 and $U_{(\psi,\psi_{p},\xi,\theta)}=\{\begin{array}{ccccc}
G_{(\psi,\psi_{p},\xi_{1},\xi_{2},\theta)}^{*} & G_{p(\psi_{p},\xi_{2})} & J_{1(\xi_{1},\psi_{p})} & J_{2(\xi_{2})} & J_{\mathrm{trt}(\theta)}\}^{T}\end{array}$ be a system of estimating functions stacking all  estimating functions
together, where $G_{p}$, $J_{1}$ and $J_{2}$ are defined in Remark
\ref{rmk1}, and $J_{\mathrm{trt}}$ is defined in (\ref{eq:ee}).
Let $(\hat{\psi},\hat{\psi}_{p},\hat{\xi},\hat{\theta})$ be the solution
to  estimating equations $P_{n}U_{(\psi,\psi_{p},\xi,\theta)}=0$.
The true parameter values are $\psi_{0}$, $\xi_{0}$ and $\theta_{0}$.
Under the regularity conditions (C1)--(C2) specified in the Supplementary
Material, if the treatment effect model $\gamma_{m,\psi}^{k}(\bar{L}_{m})$
is well specified, and either $E_{\xi_{1}}\{H_{\psi}(k)|\bar{L}_{m},\bar{A}_{m-1}=\bar{0}\}$
or $p_{\theta}(m)$ is well specified, $\hat{\psi}-\psi_{0}\rightarrow0$
in probability, as $n\rightarrow\infty$.

\end{theorem}

\begin{theorem}[Asymptotic normality] Under the regularity conditions
(C1)--(C5) specified in the Supplementary Material, $n^{1/2}(\hat{\psi}-\psi_{0})\rightarrow N_{p}\left(0,\Sigma_{\psi}\right)$
in distribution, as $n\rightarrow\infty$, where $p$ is the dimension
of $\psi_{0}$ and $\Sigma_{\psi}$ is the $p\times p$ upper left
matrix in $\{P\partial/\partial(\psi,\psi_{p},\xi,\theta)U\}^{-1}P(U$
$U^{T})\{P\partial/\partial(\psi,\psi_{p},\xi,\theta)U\}^{-1T}$. 

\end{theorem}

\begin{remark} In the statistics literature, estimators solving unbiased
estimating equations are often called $Z$-estimators. The theory
of consistency and asymptotic normality of Z-estimators is well established,
see for example Theorem 5.9 and Section 5.3 in \citet{van2000asymptotic}.
We skip the detailed proof but explain the regularity conditions needed
to guarantee the consistency in \textcolor{black}{the Supplementary
Material. From Theorem \ref{Thm1}, the functional form of $\gamma_{m,\psi}^{k}$
must be correctly specified. }In contrast, the estimator remains consistent
for $\psi$ if either $E_{\xi_{1}}\{H_{\psi}(k)|\bar{L}_{m},\bar{A}_{m-1}=\bar{0}\}$
or $p_{\theta}(m)$ is well specified, but not necessary both. The
estimator is doubly robust (\citealp{robins2001inference}; \citealp{van2003unified}).\textcolor{black}{{}
The functional form of the nuisance models can be selected on the
basis of the observed data, as well as the literature and subject
knowledge specific to the application setting. Later in this article,
we provide a more specific illustration in the context of our example. }

\end{remark}

\section{Goodness-of-fit test\label{sec:Goodness-of-Fit-Test}}

The consistency, asymptotic normality, and double robustness of estimators
rely on a key assumption, that is, the treatment effect model is well
specified. Misspecification of the treatment effect model causes bias
in the parameter estimation and break down the asymptotic results.
We now develop tests for model specification based on overidentification
restrictions tests. Conceptually, for a well specified model, a new
set of unbiased estimating functions, other than the optimal ones
that are used for estimation, evaluated at the optimal estimators,
should be asymptotically concentrated at zero. This asymptotic behavior
leads to the following theorem\textcolor{black}{: }

\begin{theorem}[Goodness-of-Fit Test] Let the treatment effect model
be $\gamma_{m,\psi}^{k}(\bar{l}_{m})$ and $H_{\psi}(k)=Y_{k}-\gamma_{T,\psi}^{k}(\bar{l}_{T})$.
Choose a set of functions $\{\tilde{q}_{m}^{k}(\bar{L}_{m})\in\mathbb{R}^{\nu},k=12,\ldots,K+1,m=k-12,\ldots,k-1\}$
that are different from the optimal choice $q_{m}^{k,\mathrm{opt}}$.
Let 
\[
\tilde{G}_{(\psi,\psi_{p},\xi,\theta)}=\sum_{k=12}^{K+1}\sum_{m=k-12}^{k-1}\tilde{q}_{m,\xi_{2}}^{k}(\bar{L}_{m})[H_{\psi}(k)-E_{\xi_{1}}\{H_{\psi_{p}}(k)\mid\bar{L}_{m},\bar{A}_{m-1}=\bar{0}\}]\{A_{m}-p_{\theta}(m)\}.
\]
Let $(\hat{\psi},\hat{\psi}_{p},\hat{\xi},\hat{\theta})$ be as in
Theorem \ref{Thm1}. The null hypothesis is $H_{0}$: $\gamma_{m}^{k}(\bar{l}_{m})$
is well specified, and either $E_{\xi_{1}}\{H_{\psi}(k)\mid\bar{L}_{m},\bar{A}_{m-1}=\bar{0}\}$
or $p_{\theta}(m)$ is well specified. Under $H_{0}$ and the regularity
conditions (C1)--(C10) specified in the Supplementary Material, the
goodness-of-fit test statistic 
\begin{equation}
GOF=n\{P_{n}\tilde{G}_{(\hat{\psi},\hat{\psi}_{p},\hat{\xi},\hat{\theta})}\}^{T}\hat{\Sigma}^{-1}P_{n}\tilde{G}_{(\hat{\psi},\hat{\psi}_{p},\hat{\xi},\hat{\theta})}\rightarrow\chi^{2}(\nu),\label{eq:chisq}
\end{equation}
in distribution, as $n\rightarrow\infty$, where $\Sigma$ is the
variance of $\Phi_{(\psi_{0},\psi_{0},\xi_{0},\theta_{0})}$ with
$\Phi_{(\psi_{0},\psi_{0},\xi_{0},\theta_{0})}$ the asymptotic linear
representation of $\tilde{G}_{(\hat{\psi},\hat{\psi}_{p},\hat{\xi},\hat{\theta})}$,
which is a linear combination of $G^{*}$, $\tilde{G}$, $G_{p}$,
$J_{1}$, $J_{2}$ and $J_{\mathrm{trt}}$, defined in (6) in\textcolor{black}{{}
the Supplementary Material,} and $\hat{\Sigma}$ is the sample variance
of $\Phi_{(\hat{\psi},\hat{\psi}_{p},\hat{\xi},\hat{\theta})}$.

\end{theorem}

We state the key steps in the proof, with details in \textcolor{black}{the
Supplementary Material.} To establish the asymptotic distribution
of $n^{1/2}P_{n}\tilde{G}_{(\hat{\psi},\hat{\psi}_{p},\hat{\xi},\hat{\theta})}$,
and therefore that of $GOF$, a key step is to linearise $n^{1/2}P_{n}\tilde{G}_{(\hat{\psi},\hat{\psi}_{p},\hat{\xi},\hat{\theta})}$
as $n^{1/2}P_{n}\Phi_{(\psi_{0},\psi_{0},\xi_{0},\theta_{0})}$ for
some $\Phi$, whereby we can apply the typical central limit theorem.
To do so, the Lipschitz condition (C7) implies the functions $\tilde{G}_{(\psi,\psi_{p},\xi,\theta)}$
form a Donsker class. Using Lemma 19.24 of \citet{van2000asymptotic},
we have 
\begin{eqnarray}
\sqrt{n}(P_{n}-P)\tilde{G}_{(\hat{\psi},\hat{\psi}_{p},\hat{\xi},\hat{\theta})} & = & \sqrt{n}(P_{n}-P)\tilde{G}_{(\psi_{0},\psi_{0},\xi_{0},\theta_{0})}+o_{p}(1).\label{eq:donsker}
\end{eqnarray}
Next, we apply a Taylor expansion to $P\tilde{G}_{(\hat{\psi},\hat{\psi}_{p},\hat{\xi},\hat{\theta})}$
on the left hand side of (\ref{eq:donsker}) and use the fact that
$P\tilde{G}_{(\psi_{0},\psi_{0},\xi_{0},\theta_{0})}=0$ on the right
hand side of (\ref{eq:donsker}). Finally, we can express $\Phi_{(\psi_{0},\psi_{0},\xi_{0},\theta_{0})}$,
the asymptotic linear representation of $\tilde{G}_{(\hat{\psi},\hat{\psi}_{p},\hat{\xi},\hat{\theta})}$,
to be a linear combination of $G^{*}$, $\tilde{G}$, $G_{p}$, $J_{1}$,
$J_{2}$ and $J_{\mathrm{trt}}$. 

\begin{remark}[Double Robustness] The goodness-of-fit test statistic
is doubly robust in the sense that for (\ref{eq:chisq}) to hold we
only require that either $E_{\xi_{1}}\{H_{\psi}(k)|\bar{L}_{m},\bar{A}_{m-1}=\bar{0}\}$
or $p_{\theta}(m)$ is well specified, not necessary for both. This
property adds a protection from possible misspecification of the nuisance
models. 

\end{remark}

\begin{remark}The standard overidentification restrictions test is
$\min_{\psi}nP_{n}V_{(\psi)}^{T}\{\hat{\Sigma}(\psi)\}^{-1}P_{n}V_{(\psi)},$
where $V_{(\psi)}\equiv(\begin{array}{cc}
G_{(\psi,\hat{\psi}_{p},\hat{\xi}_{1},\hat{\xi}_{2},\hat{\theta})}^{*} & \tilde{G}_{(\psi,\hat{\psi}_{p},\hat{\xi},\hat{\theta})}\end{array})^{T}$ and $\Sigma(\psi)$ is the asymptotic variance of $P_{n}V_{(\psi)}$.
In most situations, the minimum is obtained by a continuous iterative
procedure to update the parameter estimates; that is, $\hat{\psi}^{(t+1)}=\arg\min_{\psi}nP_{n}V_{(\psi)}^{T}\{\hat{\Sigma}(\hat{\psi}^{(t)})\}^{-1}P_{n}V_{(\psi)}$
until convergence \citep{hansen1996finite}. Our test procedure does
not need any iterative procedure, which simplifies the calculation. 

\end{remark}

\begin{remark}[Choosing $\tilde{q}$]\textcolor{black}{Just like a
naive choice of $q$ in estimating equations may lead to an estimator
with large variance and thus useless inference, an arbitrary choice
of $\tilde{q}$ may lead to the goodness-of-fit test lacking of power.
We propose the following procedure to choose $\tilde{q}$, which is
powerful in certain circumstances. Suppose we have two models to choose
from for the treatment effect. Let the null model be $\gamma_{\psi}^{*}$,
which is the treatment effect model we are testing for, and the other
model to be an alternative model $\tilde{\gamma}_{\psi}$. We can
derive $\Delta^{*}$, $q^{*\mathrm{opt}}$, $\tilde{\Delta}$, and
$\tilde{q}^{\mathrm{opt}}$ as in (\ref{eq:opt}) with $\gamma_{\psi}^{*}$
and $\tilde{\gamma}_{\psi}$. Note that $q^{*\mathrm{opt}}$ is used
for optimal estimation of the parameters in the null model. Then,
candidates for $\tilde{q}$ are $\Delta^{*}$, $\tilde{\Delta}$,
$\tilde{q}^{\mathrm{opt}}$, or any subvector of these that is not
included in $q^{*\mathrm{opt}}$. Our simulation study shows that
the goodness-of-fit test with $\tilde{q}^{\mathrm{opt}}$ is most
powerful among this set of candidates in detecting the alternative
model.}

\end{remark}

\section{Extension of goodness-of-fit test in the presence of censoring\label{sec:Extension-of-GOF-censoring}}

We use the Inverse-Probability-of-Censoring-Weighting technique (\citealp{robins1995analysis};
\citealp{hernan2005structural}; \citealp{lok2012impact}) to accommodate
patients lost to follow-up. Let $C_{p}=0$ indicate a patient remains
in the study at month $p$. Following \citet{lok2012impact}, we assume
that censoring is missing at random; that is, $(\bar{L},\bar{A})\amalg C_{k+1}\mid\bar{L}_{k},\bar{A}_{k},\bar{C}_{k}=\bar{0}$,
whereby we have $P(A_{m}=1\mid\bar{L}_{m},\bar{A}_{m-1}=\bar{0},\bar{C}_{m}=\bar{0})=P(A_{m}=1\mid\bar{L}_{m},\bar{A}_{m-1}=\bar{0})$,
and $p_{\theta}(m)$ does not depend on censoring. Define the Inverse-Probability-of-Censoring-Weightingg
version of estimating functions $G^{*c}$ and $\tilde{G}^{c}$ using
weights $W_{m,\eta}^{k}=1/\{\prod_{p=m+1}^{k}P_{\eta}(C_{p}=0\mid\bar{L}_{p-1},\bar{A}_{p-1},\bar{C}_{p-1}=\bar{0})\}$,
see the Supplementary Material for details. In calculation of the
weights, we use a pooled logistic regression model to estimate $P_{\eta}(C_{p}=0\mid\bar{L}_{p-1},\bar{A}_{p-1},\bar{C}_{p-1}=\bar{0})$.
We assume the censoring model is well specified with estimating functions
$J_{\mathrm{cen}(\eta)}$. Similarly, we have the Inverse-Probability-of-Censoring-Weighting
version of the estimating function for the preliminary estimator $\hat{\psi}_{p}$,
denoted by $G_{p}^{c}$. For the nuisance regression outcome models,
the regression was restricted to patients still in follow-up and use
weighted regression analysis with the censoring weights. 

Define the goodness-of-fit test statistic as 
\[
GOF^{c}=n\{P_{n}\tilde{G}_{(\hat{\psi},\hat{\psi}_{p},\hat{\xi},\hat{\theta},\hat{\eta})}^{c}\}^{T}(\hat{\Sigma}^{c})^{-1}P_{n}\tilde{G}_{(\hat{\psi},\hat{\psi}_{p},\hat{\xi},\hat{\theta},\hat{\eta})}^{c},
\]
where $\hat{\Sigma}^{c}$ is the sample variance of $\Phi_{(\hat{\psi},\hat{\psi}_{p},\hat{\xi},\hat{\theta},\hat{\eta})}^{c}$,
with $\Phi_{(\psi_{0},\psi_{0},\xi_{0},\theta_{0},\eta_{0})}^{c}$
the asymptotic linear representation of $\tilde{G}_{(\hat{\psi},\hat{\psi}_{p},\hat{\xi},\hat{\theta},\hat{\eta})}^{c}$,
defined by \textcolor{black}{(10)} in \textcolor{black}{the Supplementary
Material}. As proved in the Supplementary Material, subject to regularity
conditions, $GOF^{c}$ has an asymptotic chi-squared distribution,
with degrees of freedom dimension of $\tilde{G}^{c}$.

\section{Simulations\label{sec:Simulations}}

The simulation designs were based on the Acute Infection and Early
Disease Research Program database, but we did not consider censoring.
Following an unpublished 2014 technical report available from the
second author, we first generated the CD4 count outcomes $Y_{k}^{(\infty)}$
under no treatment, followed by a treatment initiation time $T$,
and lastly the observed outcomes $Y_{k}$ ($k=6,\ldots,30$), as follows:
(i) In each sample, $2$ groups were simulated: injection drug users
($10\%$) and patients who never used drugs ($90\%$), and then $\log Y_{6}^{(\infty)}\sim N(6\cdot0,0\cdot4^{2})$
for injection drug users, and $N(6\cdot6,0\cdot5^{2})$ for non injection
drug users. For $k\geq6$, $Y_{k+1}^{(\infty)}=-10+Y_{k}^{(\infty)}+\epsilon_{k+1},$
where $\epsilon_{k}\sim N(0,\sigma_{k}^{2})$ with $\sigma_{k}=52\cdot375-1\cdot625k$
for $k=7,\ldots,19$ and $\sigma_{k}=21\cdot5$ for $k=20,\ldots,30$;
(ii) $T$ was generated by a logistic regression model $\mathrm{logit}\{P(T=m\mid T\geq m,\bar{L}_{m})\}=-2\cdot4-0\cdot42\mathrm{injdrug}-0\cdot0035Y_{m}^{(\infty)}-0\cdot026m$,
where $\mathrm{injdrug}$ is an indication of being an injection drug
user; and (iii) $Y_{k}=Y_{k}^{(\infty)}+\gamma_{T}^{k}(\bar{L}_{T})$.
We considered different models for $\gamma_{m}^{k}$. 

The performance of the test statistics was assessed by their ability
(i) to confirm the adequacy of a model that is correctly specified
with the data-generating model (type-I error) and (ii) to reject a
misspecified model (power). The model under the null hypothesis $H_{0}$
upon which the goodness-of-fit statistic is based, and the alternative
hypothesis $H_{a}$ were specified as $H_{0}:\gamma_{m,\psi}^{k}=(\psi_{1}+\psi_{2}m)(k-m)$
versus $H_{a}:\gamma_{m,\psi}^{k}\neq(\psi_{1}+\psi_{2}m)(k-m).$
Six scenarios regarding the true treatment effect model, $H_{0}$
and a parametric specification of $H_{a}$ were specified as follows:

Scenario (a): True: $\gamma_{m,\psi}^{k}=(25-0\cdot7m)(k-m)$, $H_{0}:$
$\gamma_{m,\psi}^{k}=(\psi_{1}+\psi_{2}m)(k-m)$, and $H_{a}:$ $\gamma_{m,\psi}^{k}=(\psi_{1}+\psi_{2}m+\psi_{3}m^{2})(k-m)$; 

Scenario (b): True: $\gamma_{m,\psi}^{k}=(25-0\cdot7m)(k-m)$, $H_{0}:$
$\gamma_{m,\psi}^{k}=(\psi_{1}+\psi_{2}m+\psi_{3}\mathrm{injdrug})(k-m)$,
and $H_{a}:$ $\gamma_{m,\psi}^{k}=(\psi_{1}+\psi_{2}m+\psi_{3}m^{2})(k-m)$; 

Scenario (c): True: $\gamma_{m,\psi}^{k}=(35-1\cdot1m+0\cdot04m^{2})(k-m)$,
$H_{0}:$ $\gamma_{m,\psi}^{k}=(\psi_{1}+\psi_{2}m)(k-m)$, and $H_{a}:$
$\gamma_{m,\psi}^{k}=(\psi_{1}+\psi_{2}m+\psi_{3}m^{2})(k-m)$; 

Scenario (d): True: $\gamma_{m,\psi}^{k}=(35-1\cdot1m+0\cdot04k^{2})(k-m)$,
$H_{0}:$ $\gamma_{m,\psi}^{k}=(\psi_{1}+\psi_{2}m)(k-m),$ and $H_{a}:$
$\gamma_{m,\psi}^{k}=(\psi_{1}+\psi_{2}m+\psi_{3}m^{2})(k-m)$; 

Scenario (e): True: $\gamma_{m,\psi}^{k}=(25-m+0\cdot03m^{2})(k-m)$,
$H_{0}:$ $\gamma_{m,\psi}^{k}=(\psi_{1}+\psi_{2}m)(k-m)$, and $H_{a}:$
$\gamma_{m,\psi}^{k}=(\psi_{3}+\psi_{4}m)(k-m)^{3/2}$; 

Scenario (f): True: $\gamma_{m,\psi}^{k}=(10-1\cdot1m)(k-m)^{3/2}$,
$H_{0}:$ $\gamma_{m,\psi}^{k}=(\psi_{1}+\psi_{2}m)(k-m),$ and $H_{a}:$
$\gamma_{m,\psi}^{k}=(\psi_{3}+\psi_{4}m)(k-m)^{3/2}$. 

Specifically, in Scenarios (a) and (b), $H_{0}$ is correctly specified.
In Scenarios (c)--(f), $H_{0}$ is misspecified with different degrees
of departure from the true model. In Scenarios (c) and (d), $H_{0}$
is nested in the parametric specification of $H_{a}$. In Scenarios
(e) and (f), $H_{0}$ is not nested in the parametric specification
of $H_{a}$. 

Type-I error and power were estimated by the frequency of rejecting
$H_{0}$ using $1,000$ simulated datasets.\textcolor{black}{{} We considered
the following choices of $\tilde{q}$: (i) $\tilde{q}_{m}^{k}\equiv1$,
which is a naive choice for comparison; }(ii)\textcolor{black}{{} $\tilde{q}_{m}^{k}=\tilde{\Delta}_{m}^{k}$;
and (iii) $\tilde{q}_{m}^{k}=\tilde{q}_{m}^{\mathrm{opt},k}$. (ii)
and (iii) were derived under the parametric specification of $H_{a}$. }

The optimal estimator was obtained by solving (\ref{eq:ee}) with
(\ref{eq:opt}). In (\ref{eq:ee}), the treatment initiation model
was fitted by a logistic regression model adjusting for $\mathrm{Y}_{m}$,
injection drug use, and month, restricted to patients and visits with
$\bar{A}_{m-1}=\bar{0}$. Thus, the treatment initiation model was
correctly specified. $E\{H_{\psi_{p}}(k)\mid\bar{A}_{m-1}=\bar{0},\bar{L}_{m}\}$
was fitted by a linear regression model adjusting for $\mathrm{CD4}_{m}$
and $(k-m)$, restricted to patients and visits with $\bar{A}_{m-1}=\bar{0}$.
The covariates were motivated by (\ref{eq:imp1ofNUC}) and the data
generating mechanism. The nuisance models in $q_{m}^{\mathrm{opt}}$
are specified in the Supplementary Material  for simplicity of presentation,
which do not affect the double robustness of the estimator. 

\textcolor{black}{In addition to the goodness-of-fit test statistic,
we considered an elaborated-model-fitting-and-testing approach, which
combines the null model and the parametric specification of the alternative
model and tests the significance of the parameters corresponding to
the alternative model. }

\textcolor{black}{From Scenarios (a) and (b) in Table \ref{tab:1},
where the treatment effect model under $H_{0}$ is correctly specified,
the goodness-of-fit test procedure with all choices of $\tilde{q}$
controls type-I error for $n=1,000$ and $n=2,000$. This suggests
that the chi-squared distribution derived in this article provides
an accurate approximation to the finite sample behavior of the goodness-of-fit
test statistic for these sample sizes.}

\textcolor{black}{From Scenarios (c)--(f), where the treatment effect
model is not correctly specified, the goodness-of-fit test procedure
with the optimal $\tilde{q}_{m}^{k}$ derived under the parametric
specification of $H_{a}$ is most powerful, and as the sample size
increases, the power increases, confirming the theoretical results.
From Scenarios (c) and (d), the goodness-of-fit test procedure and
the elaborated-model-fitting-and-testing approach are comparable when
testing nested models. In both scenarios, the goodness-of-fit test
procedure is slightly more powerful than the elaborated-model-fitting-and-testing
approach for $n=500$ and $n=1,000$, which is not apparent for $n=2,000$.
For Scenarios (e) and (f), the null treatment effect model is not
nested in the parametric specification of $H_{a}$. Under Scenario
(e),} the goodness-of-fit test statistic with \textcolor{black}{$\tilde{q}_{m}^{\mathrm{opt},k}$
shows more power than the elaborated-model-fitting-and-testing approach,
likely because the elaborated-model-fitting-and-testing approach fits
a larger model and loses power. Under Scenario (f), the goodness-of-fit
test statistic with $\tilde{q}_{3m}^{\mathrm{opt},k}$ is slightly
more powerful than the elaborated-model-fitting-and-testing approach
for $n=500$, and both approaches are powerful to reject the null
model in other cases.}

\begin{table}[H]
\centering{}{\scriptsize{}{}\protect\caption{Type-I error estimates and power estimates ($\times100$) for testing
the null model $H_{0}$ by the proposed goodness-of-fit (GOF) test
statistic with $\tilde{q}$ being $1$, $\tilde{\Delta}$, and $\tilde{q}^{\mathrm{opt}}$,
and the Elaborated Model Fitting and Testing (EMFT) approach over
$1,000$ simulations under Scenarios (a)--(f) \label{tab:1}}
}{\scriptsize \par}

{\scriptsize{}{}}%
\begin{tabular}{ccccccccc}
 & \multicolumn{4}{c}{Type-I error estimates in Scenario (a)} & \multicolumn{4}{c}{Type-I error estimates in Scenario (b)}\tabularnewline
 &  & GOF &  & EMFT &  & GOF &  & EMFT\tabularnewline
$n$\textbackslash{}$\tilde{q}$  & $1$  & $\tilde{\Delta}_{m}^{k}$  & $\tilde{q}_{m}^{\mathrm{opt},k}$ &  & $1$  & $\tilde{\Delta}_{3m}^{k}$  & $\tilde{q}_{m}^{\mathrm{opt},k}$ & \tabularnewline
$500$  & $5\cdot3$ & $4\cdot3$ & $5\cdot2$ & $4\cdot9$ & \textcolor{black}{$9\cdot1$} & \textcolor{black}{$9\cdot8$} & \textcolor{black}{$12\cdot3$} & \textcolor{black}{$13\cdot5$}\tabularnewline
$1000$  & $4\cdot5$ & $5\cdot7$ & $5\cdot6$ & $5\cdot4$ & \textcolor{black}{$5\cdot3$} & \textcolor{black}{$4\cdot4$} & \textcolor{black}{$5\cdot4$} & \textcolor{black}{$5\cdot7$}\tabularnewline
$2000$  & $4\cdot8$ & $4\cdot4$ & $5\cdot2$ & $5\cdot3$ & \textcolor{black}{$5\cdot2$} & \textcolor{black}{$4\cdot4$} & \textcolor{black}{$5\cdot2$} & \textcolor{black}{$5\cdot3$}\tabularnewline
 & \multicolumn{4}{c}{Power estimates in Scenario (c)} & \multicolumn{4}{c}{Power estimates in Scenario (d)}\tabularnewline
 &  & GOF &  & EMFT &  & GOF &  & EMFT\tabularnewline
$n$\textbackslash{}$\tilde{q}$  & $1$  & $\tilde{\Delta}_{m}^{k}$  & $\tilde{q}_{m}^{\mathrm{opt},k}$ &  & $1$  & $\tilde{\Delta}_{m}^{k}$  & $\tilde{q}_{m}^{\mathrm{opt},k}$ & \tabularnewline
$500$  & $15$ & $29$ & $59$ & $56$ & $90$ & $96$ & $97$ & $83$\tabularnewline
$1000$  & $28$ & $55$ & $89$ & $84$ & $100$ & $100$ & $100$ & $97$\tabularnewline
$2000$  & $52$ & $88$ & $99$ & $99$ & $100$ & $100$ & $100$ & $100$\tabularnewline
 & \multicolumn{4}{c}{Power estimates in Scenario (e)} & \multicolumn{4}{c}{Power estimates in Scenario (f)}\tabularnewline
 &  & GOF &  & EMFT &  & GOF &  & EMFT\tabularnewline
$n$\textbackslash{}$\tilde{q}$  & $1$  & $\tilde{\Delta}_{m}^{k}$  & $\tilde{q}_{m}^{\mathrm{opt},k}$ &  & $1$  & $\tilde{\Delta}_{m}^{k}$  & $\tilde{q}_{m}^{\mathrm{opt},k}$ & \tabularnewline
$500$  & $12$ & $25$ & $49$ & $28$ & $93$ & $99$ & $100$ & $96$\tabularnewline
$1000$  & $24$ & $53$ & $73$ & $54$ & $100$ & $100$ & $100$ & $100$\tabularnewline
$2000$  & $48$ & $79$ & $91$ & $80$ & $100$ & $100$ & $100$ & $100$\tabularnewline
\end{tabular}
\end{table}

\section{Application\label{sec:Application-to-Initiating}}

We applied the proposed goodness-of-fit test to study how the timing
of combination antiretroviral treatment initiation after infection
predicts the effect of one year of treatment in HIV-positive patients.
We used the Acute Infection Early Disease Research Program database.
We started with a simple null model for the treatment effect, \textcolor{black}{$H_{0}:\gamma_{m,\psi}^{k}=(\psi_{1}+\psi_{2}m)(k-m)$},
and conducted directed alternative-model tests by testing whether
possible effect modifiers should be added into the model. In the HIV
literature, it has been found that there may be gender differences
in immunologic response to combination antiretroviral treatment: early
studies suggested that clinical disease progression was more rapid
in women than men with combination antiretroviral treatment \citep{friedland1991survival,bozzette1998care};
conversely, more recent studies have shown that women have better
immunologic outcomes than men on treatment\citep{maman2012gender,maskew2013gender}.
It has also been shown that older age is associated with a poorer
CD4 count increase with combination antiretroviral treatment \citep{maman2012gender,maskew2013gender}.
Injection drug use has been found to be associated with reduced effectiveness
of combination antiretroviral treatment \citep{poundstone2001differences}.
As suggested by the literature, we considered tests directed at $3$
variables: gender, age, and injection drug use. \textcolor{black}{For
the test directed at a certain variable $Z$, we calculated the goodness-of-fit
test statistic with $\tilde{q}$ being the optimal form derived from
the parametric specification of the alternative model $\tilde{\gamma}_{m,\psi}^{k}=(\psi_{1}+\psi_{2}m+\psi_{3}Z)(k-m)1_{(k>m)}$. }

The nuisance models were specified on the basis of the observed data,
the clinical literature, and subject knowledge. For the censoring
model, we used a logistic regression model adjusting for square root
of current CD4 count ($\mathrm{CD4}_{m}^{1/2}$), current log viral
load, gender, age, \textcolor{black}{injection drug use} (injdrug),
month, squared month, and whether a patient was treated, as discussed
in \citet{krishnan2011incidence} and \citet{lok2010long}. For the
treatment initiation model, we used a logistic regression models including
$\mathrm{CD4}_{m}^{1/2}$, current log viral load, gender, age, injection
drug use, month, days since last visit, indication of first visit,
indication of second visit, race, as discussed in \citet{lok2014opt}.
For $E_{\xi_{2}}\{H(k)\mid\bar{L}_{m},\bar{A}_{m}=\bar{0}\}$, we
used a regression model adjusting for $\mathrm{CD4}_{m}$, $\mathrm{CD4}_{m}^{3/4}(k-m)$,
$\mathrm{CD4}_{m}^{3/4}\mathrm{age}(k-m)$, $\mathrm{CD4}_{m}^{3/4}\mathrm{race}(k-m)$,
$\mathrm{CD4}_{m}^{3/4}\mathrm{injdrug}(k-m)$, whether there is a
CD4 slope measure, $\mathrm{CD4slope}_{m}(k-m)^{1/2}$, $(6-m)^{+}$,
and $(6^{2}-m^{2})^{+}$ with $a^{+}\equiv a\times1(a>0)$. The model
was motivated by (\ref{eq:imp1ofNUC}). The inclusion of $\mathrm{CD4}_{m}$,
$\mathrm{CD4}_{m}^{3/4}(k-m)$, and $\mathrm{CD4}_{m}^{3/4}\mathrm{age}(k-m)$
was suggested from a stochastic model of $Y_{ik}^{(\infty)}$ for
each patient $i$ over time $k$, $\{Y_{ik}^{(\infty)}\}^{1/4}=a_{i}+bk+\gamma_{1}\mathrm{age}+\gamma_{2}\mathrm{age}k+\phi W_{ik}+\epsilon_{ik},$
where $a_{i}$ is a normal random effect, $W_{ik}$ is a Brownian
motion process, $\epsilon_{ik}$ is normal with mean zero and constant
variance, and $a_{i}$, $W_{ik}$, $\epsilon_{ik}$ and $\mathrm{age}$
are independent \citep{taylor1994stochastic}. Other covariates were
suggested in \citet{taylor1998does} and \citet{may2009cd4}.

Table \ref{tab:The-AIEDRP-data:} shows the results from fitting the
optimal estimator of the null treatment effect model, along with the
goodness-of-fit tests directed at gender, age, and injection drug
use. The $p$-values are all greater than $0\cdot05$. To avoid the
multiple testing problem, we did not consider other tests. The three
tests were specified prior to the actual calculation. \textcolor{black}{The
results show a benefit of combination antiretroviral treatment; for
example, starting treatment at the estimated date of infection would
lead to an expected added improvement in CD4 counts of $12\hat{\psi}_{1}=299$
}$\text{cells/mm}^{3}$\textcolor{black}{{} after a year of therapy.
Delaying treatment initiation during acute and early infection may
diminish the CD4 count gain associated with one year treatment ($\hat{\psi}_{2}<0$);
however, this result is not statistically significant. }

\begin{table}[H]
\centering{}{\scriptsize{}{}\protect\caption{\label{tab:The-AIEDRP-data:}The Acute Infection Early Disease Research
Program data: the optimal estimator fitting the null treatment effect
model: point estimate ($95\%$ confidence intervals based on the asymptotic
normality result), along with goodness-of-fit statistics (Statistic),
associated degree of freedom (DF), and $p$-values ($p$-value) for
the adequacy of the null model by testing whether gender or injection
drug use should be added into the model}
}{\scriptsize \par}

\begin{tabular}{cccc}
\multicolumn{2}{c}{$\hat{\psi}_{1}$($95\%$ CI)} & \multicolumn{2}{c}{$\hat{\psi}_{2}$ ($95\%$ CI)}\tabularnewline
\multicolumn{2}{c}{$24\cdot88(21\cdot61,28\cdot15)$} & \multicolumn{2}{c}{$-0\cdot48(-1\cdot47,0\cdot52)$}\tabularnewline
\multicolumn{4}{c}{Goodness-of-fit test}\tabularnewline
 & Statistic & DF & $p$-value\tabularnewline
Test directed at gender & $0\cdot99$ & $1$ & $0\cdot32$\tabularnewline
Test directed at age & $0\cdot80$ & $1$ & $0\cdot37$\tabularnewline
Test directed at injection drug use & $2\cdot93$ & $1$ & $0\cdot09$\tabularnewline
\end{tabular}
\end{table}

\section{Discussion\label{sec:Discussion}}

The applicability of the goodness-of-fit test procedure presented
in this article is broad in the causal inference literature. The testing
procedure can also be developed for the traditional structural nested
mean models \citep{robins1994correcting} other than the time-dependent
coarse structural nested mean models considered in this article, and
marginal structural models \citep{robins2000marginal}, because both
approaches yield overidentification of the parameters. For the Inverse-Probability-of-Censoring-Weighting
estimator of marginal structural models, unbiased estimating equations
are $P_{n}q(V)\{Y-\mu(\bar{A},V)\}w(\bar{A}\mid\bar{L})=0,$ where
$Y$ is the outcome at the end of study, $\bar{A}$ is the treatment
history, $V$ is a subset of the baseline covariates, $\mu(\bar{a},V)\equiv E(Y^{\bar{a}}\mid V)$
is the marginal structural model, where $Y^{\bar{a}}$ is the counterfactual
outcome had every individual received the treatment $\bar{a}$, and
$w(\bar{A}\mid\bar{L})$ is the inverse of conditional probability
of receiving the actual treatment $\bar{A}$ given $\bar{L}$. These
equations are unbiased for most choices of $q(V)$, leading to a large
class of unbiased estimating equations. The literature of structural
nested mean models and marginal structural models concentrates on
estimation and efficiency. Little attention has been given to goodness-of-fit
tests. Our test procedure for the treatment effect model can be developed
in these contexts in the same manner. 

Our goodness-of-fit test procedure can also deal with treatment of
the form ``initiate treatment when the CD4 count first drops below
$x$''. Especially in resource limited countries, and historically
also in the US, initiation of combination antiretroviral treatment
is decided based on the CD4 count threshold. \citet{orellana2010dynamic_a}
proposed dynamic regime marginal structural models and \citet{lok2007optimalstart}
used structural nested mean models to simultaneously compare dynamic
treatment regimes of this form and estimate the optimal one. Due to
the popularity of these methods, the development of goodness-of-fit
tests in these settings will be useful for model diagnosis and protect
causal estimates from biases introduced by misspecification of the
treatment effect model.

\section*{Acknowledgements}

We are grateful to the patients who volunteered for AIEDRP, to the
AIEDRP study team, and to Susan Little, Davey Smith, and Christy Anderson
for their help and advice in interpreting the AIEDRP database. We
would like to thank James Robins and Victor DeGruttola for insightful
and fruitful discussions. This work was sponsored by the Milton Fund,
the Career Incubator Fund award from the Harvard School of Public
Health, and NIH grants NIAID R01AI100762, R01 AI51164, R37 AI032475,
AI43638, AI74621, and AI36214.

\section*{Supplementary material}

Supplementary material available at \textit{Biometrika} online includes
regularity conditions, the proof of Theorem 3, the derivation of the
goodness-of-fit test statistic in the presence of censoring, and the
nuisance regression outcome models used in the simulation.



\bibliographystyle{plainnat}
\bibliography{mybib2}

\begin{thebibliography}{36}
\providecommand{\natexlab}[1]{#1}
\providecommand{\url}[1]{\texttt{#1}}
\expandafter\ifx\csname urlstyle\endcsname\relax
  \providecommand{\doi}[1]{doi: #1}\else
  \providecommand{\doi}{doi: \begingroup \urlstyle{rm}\Url}\fi

\bibitem[Arellano and Bond(1991)]{arellano1991some}
Manuel Arellano and Stephen Bond.
\newblock Some tests of specification for panel data: Monte carlo evidence and
  an application to employment equations.
\newblock \emph{Rev. econ. stud.}, 58\penalty0 (2):\penalty0 277--297, 1991.

\bibitem[Bozzette et~al.(1998)Bozzette, Berry, Duan, Frankel, Leibowitz,
  Lefkowitz, Emmons, Senterfitt, Berk, Morton, et~al.]{bozzette1998care}
Samuel~A Bozzette, Sandra~H Berry, Naihua Duan, Martin~R Frankel, Arleen~A
  Leibowitz, Doris Lefkowitz, Carol-Ann Emmons, J~Walton Senterfitt, Marc~L
  Berk, Sally~C Morton, et~al.
\newblock The care of {HIV}-infected adults in the united states.
\newblock \emph{New Engl. J. Med.}, 339\penalty0 (26):\penalty0 1897--1904,
  1998.

\bibitem[Dawid(1979)]{dawid1979conditional}
A~Philip Dawid.
\newblock Conditional independence in statistical theory.
\newblock \emph{J. R. Statist. Soc. B}, 41:\penalty0 1--31, 1979.

\bibitem[Friedland et~al.(1991)Friedland, Saltzman, Vileno, Freeman, Schrager,
  and Klein]{friedland1991survival}
Gerald~H Friedland, Brian Saltzman, Joan Vileno, Katherine Freeman, Lewis~K
  Schrager, and Robert~S Klein.
\newblock Survival differences in patients with {AIDS}.
\newblock \emph{JAIDS J. Acq. Im. Deficien. Synd.}, 4\penalty0 (2):\penalty0
  144--153, 1991.

\bibitem[Hansen(1982)]{hansen1982large}
Lars~Peter Hansen.
\newblock Large sample properties of generalized method of moments estimators.
\newblock \emph{Econometrica: J. Economet. Soc.}, 50:\penalty0 1029--1054,
  1982.

\bibitem[Hansen et~al.(1996)Hansen, Heaton, and Yaron]{hansen1996finite}
Lars~Peter Hansen, John Heaton, and Amir Yaron.
\newblock Finite-sample properties of some alternative gmm estimators.
\newblock \emph{J. Busi. \& Econ. Stat.}, 14\penalty0 (3):\penalty0 262--280,
  1996.

\bibitem[Hecht et~al.(2006)Hecht, Wang, Collier, Little, Markowitz, Margolick,
  Kilby, Daar, and Conway]{hecht2006multicenter}
Frederick~M Hecht, Lei Wang, Ann Collier, Susan Little, Martin Markowitz,
  Joseph Margolick, J~Michael Kilby, Eric Daar, and Brian Conway.
\newblock A multicenter observational study of the potential benefits of
  initiating combination antiretroviral therapy during acute {HIV} infection.
\newblock \emph{J. Infect. Dis.}, 194\penalty0 (6):\penalty0 725--733, 2006.

\bibitem[Hern{\'a}n et~al.(2005)Hern{\'a}n, Cole, Margolick, Cohen, and
  Robins]{hernan2005structural}
Miguel~A Hern{\'a}n, Stephen~R Cole, Joseph Margolick, Mardge Cohen, and
  James~M Robins.
\newblock Structural accelerated failure time models for survival analysis in
  studies with time-varying treatments.
\newblock \emph{Pharm. \& Drug Saf.}, 14\penalty0 (7):\penalty0 477--491, 2005.

\bibitem[Imbens et~al.(1995)Imbens, Johnson, and Spady]{imbens1995information}
Guido Imbens, Phillip Johnson, and Richard~H Spady.
\newblock \emph{Information theoretic approaches to inference in moment
  condition models}.
\newblock National Bureau of Economic Research Cambridge, Mass., USA, 1995.

\bibitem[Krishnan et~al.(2011)Krishnan, Wu, Smurzynski, Bosch, Benson, Collier,
  Klebert, Feinberg, and Koletar]{krishnan2011incidence}
S~Krishnan, K~Wu, M~Smurzynski, RJ~Bosch, CA~Benson, AC~Collier, MK~Klebert,
  J~Feinberg, and SL~Koletar.
\newblock Incidence rate of and factors associated with loss to follow-up in a
  longitudinal cohort of antiretroviral-treated {HIV}-infected persons: an
  {AIDS Clinical Trials Group (ACTG) Longitudinal Linked Randomized Trials
  (ALLRT)} analysis.
\newblock \emph{HIV Clin. Tria.}, 12\penalty0 (4):\penalty0 190--200, 2011.

\bibitem[Lok et~al.(2007)Lok, H{\'e}rnan, and Robins]{lok2007optimalstart}
J.~J. Lok, M.~A. H{\'e}rnan, and J.~M. Robins.
\newblock Optimal start of {HAART} treatment in {HIV} positive patients.
\newblock \emph{Proceedings of the Joint Statistical Meetings}, pages
  1149--1160, 2007.

\bibitem[Lok et~al.(2004)Lok, Gill, Van Der~Vaart, and
  Robins]{lok2004estimating}
Judith Lok, Richard Gill, Aad Van Der~Vaart, and James Robins.
\newblock Estimating the causal effect of a time-varying treatment on
  time-to-event using structural nested failure time models.
\newblock \emph{Statist. Neerland.}, 58\penalty0 (3):\penalty0 271--295, 2004.

\bibitem[Lok and DeGruttola(2012)]{lok2012impact}
Judith~J Lok and Victor DeGruttola.
\newblock Impact of time to start treatment following infection with
  application to initiating haart in {HIV}-positive patients.
\newblock \emph{Biometrics}, 68\penalty0 (3):\penalty0 745--754, 2012.

\bibitem[Lok and Griner(2014)]{lok2014opt}
Judith~J Lok and Ray Griner.
\newblock Optimal estimation of coarse structural nested mean models with
  application to initiating haart in {HIV}-positive patients.
\newblock \emph{Submitted}, 2014.

\bibitem[Lok et~al.(2010)Lok, Bosch, Benson, Collier, Robbins, Shafer, and
  Hughes]{lok2010long}
Judith~J Lok, Ronald~J Bosch, Constance~A Benson, Ann~C Collier, Gregory~K
  Robbins, Robert~W Shafer, and Michael~D Hughes.
\newblock Long-term increase in {CD4+ T-cell} counts during combination
  antiretroviral therapy for {HIV}-1 infection.
\newblock \emph{AIDS (London, England)}, 24\penalty0 (12):\penalty0 1867--1876,
  2010.

\bibitem[Maman et~al.(2012)Maman, Pujades-Rodriguez, Subtil, Pinoges, McGuire,
  Ecochard, and Etard]{maman2012gender}
David Maman, Mar Pujades-Rodriguez, Fabien Subtil, Loretxu Pinoges, Megan
  McGuire, Rene Ecochard, and Jean-Fran{\c{c}}ois Etard.
\newblock Gender differences in immune reconstitution: a multicentric cohort
  analysis in {sub-Saharan Africa}.
\newblock \emph{PLoS One}, 7\penalty0 (2):\penalty0 e31078, 2012.

\bibitem[Maskew et~al.(2013)Maskew, Brennan, Westreich, McNamara, MacPhail, and
  Fox]{maskew2013gender}
Mhairi Maskew, Alana~T Brennan, Daniel Westreich, Lynne McNamara, A~Patrick
  MacPhail, and Matthew~P Fox.
\newblock Gender differences in mortality and {CD4} count response among
  virally suppressed {HIV}-positive patients.
\newblock \emph{J. Wom. Health.}, 22\penalty0 (2):\penalty0 113--120, 2013.

\bibitem[May et~al.(2009)May, Wood, Myer, Taff{\'e}, Rauch, Battegay, and
  Egger]{may2009cd4}
Margaret May, Robin Wood, Landon Myer, Patrick Taff{\'e}, Andri Rauch, Manuel
  Battegay, and Matthias Egger.
\newblock {CD4 T-cell} declines by ethnicity in untreated {HIV}-1 infected
  patients in {South Africa and Switzerland}.
\newblock \emph{J. Infect. Dis.}, 200\penalty0 (11):\penalty0 1729--1735, 2009.

\bibitem[Newey and McFadden(1994)]{newey1994large}
Whitney~K Newey and Daniel McFadden.
\newblock Large sample estimation and hypothesis testing.
\newblock \emph{Handbook economet.}, 4:\penalty0 2111--2245, 1994.

\bibitem[Orellana et~al.(2010)Orellana, Rotnitzky, and
  Robins]{orellana2010dynamic_a}
Liliana Orellana, Andrea Rotnitzky, and James~M Robins.
\newblock Dynamic regime marginal structural mean models for estimation of
  optimal dynamic treatment regimes, part i: main content.
\newblock \emph{Internat. J. Biostat.}, 6\penalty0 (2):\penalty0 1557--79,
  2010.

\bibitem[Poundstone et~al.(2001)Poundstone, Chaisson, and
  Moore]{poundstone2001differences}
Katharine~E Poundstone, Richard~E Chaisson, and Richard~D Moore.
\newblock Differences in {HIV} disease progression by injection drug use and by
  sex in the era of highly active antiretroviral therapy.
\newblock \emph{AIDS}, 15\penalty0 (9):\penalty0 1115--1123, 2001.

\bibitem[Robins(1994)]{robins1994correcting}
James~M Robins.
\newblock Correcting for non-compliance in randomized trials using structural
  nested mean models.
\newblock \emph{Comm. Stat. Theo. Meth.}, 23\penalty0 (8):\penalty0 2379--2412,
  1994.

\bibitem[Robins(1998{\natexlab{a}})]{robins1998correction}
James~M Robins.
\newblock Correction for non-compliance in equivalence trials.
\newblock \emph{Stat. in Med.}, 17\penalty0 (3):\penalty0 269--302,
  1998{\natexlab{a}}.

\bibitem[Robins(1998{\natexlab{b}})]{robins1998structural}
James~M Robins.
\newblock Structural nested failure time models.
\newblock \emph{Encycl.Biostat.}, 7, 1998{\natexlab{b}}.

\bibitem[Robins(2000)]{robins2000marginal}
James~M Robins.
\newblock Marginal structural models versus structural nested models as tools
  for causal inference.
\newblock In \emph{Statist. Mod. in Epidem., Environ., \& Clin. Trials}, pages
  95--133. Springer, 2000.

\bibitem[Robins and Rotnitzky(2001)]{robins2001inference}
James~M Robins and Andrea Rotnitzky.
\newblock Inference for semiparametric models: Some questions and an
  answer-comments.
\newblock \emph{Statist. Sinica}, 11:\penalty0 863--885, 2001.

\bibitem[Robins et~al.(1992)Robins, Blevins, Ritter, and Wulfsohn]{robins1992g}
James~M Robins, Donald Blevins, Grant Ritter, and Michael Wulfsohn.
\newblock G-estimation of the effect of prophylaxis therapy for pneumocystis
  carinii pneumonia on the survival of {AIDS} patients.
\newblock \emph{Epidemiology}, 3:\penalty0 319--336, 1992.

\bibitem[Robins et~al.(1995)Robins, Rotnitzky, and Zhao]{robins1995analysis}
James~M Robins, Andrea Rotnitzky, and Lue~Ping Zhao.
\newblock Analysis of semiparametric regression models for repeated outcomes in
  the presence of missing data.
\newblock \emph{J. Am. Statist. Assoc.}, 90\penalty0 (429):\penalty0 106--121,
  1995.

\bibitem[Robins et~al.(2000)Robins, Hernan, and
  Brumback]{robins2000marginalstructural}
James~M Robins, Miguel~Angel Hernan, and Babette Brumback.
\newblock Marginal structural models and causal inference in epidemiology.
\newblock \emph{Epidemiology}, 11\penalty0 (5):\penalty0 550--560, 2000.

\bibitem[Rubin(1978)]{rubin1978bayesian}
Donald~B Rubin.
\newblock Bayesian inference for causal effects: The role of randomization.
\newblock \emph{Ann. Statist.}, 6:\penalty0 34--58, 1978.

\bibitem[Sargan(1958)]{sargan1958estimation}
John~D Sargan.
\newblock The estimation of economic relationships using instrumental
  variables.
\newblock \emph{Econometrica: J. Economet. Soc.}, 26:\penalty0 393--415, 1958.

\bibitem[Smith et~al.(2006)Smith, Strain, Frost, Pillai, Wong, Wrin, Liu,
  Petropolous, Daar, Little, et~al.]{smith2006lack}
Davey~M Smith, Matthew~C Strain, Simon~DW Frost, Satish~K Pillai, Joseph~K
  Wong, Terri Wrin, Yang Liu, Christos~J Petropolous, Eric~S Daar, Susan~J
  Little, et~al.
\newblock Lack of neutralizing antibody response to {HIV}-1 predisposes to
  superinfection.
\newblock \emph{Virology}, 355\penalty0 (1):\penalty0 1--5, 2006.

\bibitem[Taylor and Law(1998)]{taylor1998does}
Jeremy~MG Taylor and Ngayee Law.
\newblock Does the covariance structure matter in longitudinal modelling for
  the prediction of future {CD4} counts?
\newblock \emph{Stat. in Med.}, 17\penalty0 (20):\penalty0 2381--2394, 1998.

\bibitem[Taylor et~al.(1994)Taylor, Cumberland, and Sy]{taylor1994stochastic}
Jeremy~MG Taylor, WG~Cumberland, and JP~Sy.
\newblock A stochastic model for analysis of longitudinal {AIDS} data.
\newblock \emph{J. Am. Statist. Assoc.}, 89\penalty0 (427):\penalty0 727--736,
  1994.

\bibitem[Van~der Laan and Robins(2003)]{van2003unified}
Mark~J Van~der Laan and James~M Robins.
\newblock \emph{Unified methods for censored longitudinal data and causality}.
\newblock Springer Science \& Business Media, 2003.

\bibitem[Van~der Vaart(2000)]{van2000asymptotic}
Aad~W Van~der Vaart.
\newblock \emph{Asymptotic statistics}, volume~3.
\newblock Cambridge university press, 2000.

\end{thebibliography}

\end{document}